# Quantitative Comparison between Electronic Raman Scattering and Angle-Resolved Photoemission Spectra in $Bi_2Sr_2CaCu_2O_{8+\delta}$ Superconductors: Doping Dependence of Nodal and Antinodal Superconducting Gaps

Kiyohisa Tanaka[1,2], NguyenTrung Hieu[1,3], Giulio Vincini[1], Takahiko Masui[1], Shigeki Miyasaka[1], Setsuko Tajima[1], and Takao Sasagawa[4]

[1]*Department of Physics, Osaka University, Osaka 560-0043, Japan*
[2]*UVSOR Facility, Institute for Molecular Science, Okazaki 444-8585, Japan*
[3]*Laboratory of Magnetism and Superconductivity, Institute of Materials Science, Vietnam Academy of Science and Technology, 18 Hoang Quoc Viet Str., Cau Giay, Hanoi, Vietnam*
[4]*Materials and Structures Laboratory, Tokyo Institute of Technology, Kanagawa 226-8503, Japan*

Both electronic Raman scattering (ERS) and angle-resolved photoemission spectra (ARPES) revealed two energy scales for the gap in different momentum spaces in the cuprates. However, the interpretations were different, and the gap values were also different in two experiments. In order to clarify the origin of these discrepancies, we directly compared ERS and ARPES by calculating ERS from the experimental data of ARPES through the Kubo formula. The calculated ERS spectra were in good agreement with the experimental results except for the $B_{1g}$ peak energies. The doping-dependent $B_{2g}$ peak energy was well reproduced from a doping-independent *d*-wave gap deduced from ARPES, by assuming a particular spectral weight distribution along the Fermi surface. The $B_{1g}$ peak energies could not be reproduced by the ARPES data. The difference between $B_{1g}$ ERS and antinodal ARPES became larger with underdoping, which implies that the effect of the pseudogap is different in these two techniques.

## 1. Introduction

A superconducting energy gap is a crucially important parameter for a superconductor. For high-temperature superconducting cuprates, many experimental techniques have been devoted to studying the gap energy and structure. Since a *d*-wave symmetry of the gap has been revealed[1-3], the momentum selective probes such as angle-resolved photoemission spectra (ARPES) and electronic Raman scattering (ERS) spectra have become important. Both ERS and ARPES measurements can give momentum (*k*)-selective information about the electronic structure. ARPES, a single-particle excitation method, probes the density of states at each $k$-point in the Brillouin zone (BZ), while Raman, a two-particle excitation method, probes a particular region of the BZ depending on the polarization.[4,5]

A significant problem in the physics of cuprate superconductors is that there seem to be two energy scales for a gap. The gap in the antinodal region of a BZ probed by ARPES is much larger than the value expected from the $d$-wave gap in the nodal region.[6] A further problem is that the antinodal gap increases with underdoping (namely, with decreasing $T_c$), while the nodal gap is constant with doping.[6,7] Therefore, the antinodal and nodal gaps are considered to indicate two different energy scales of the gap.

On the other hand, in ERS measurements, the spectrum with $B_{1g}$ polarization shows a pair-breaking excitation around the antinodal region, while the spectrum with $B_{2g}$ polarization indicates a gap in the nodal region. The unusual behavior of ERS is that the $B_{1g}$ pair-breaking peak energy increases monotonically with underdoping, while the $B_{2g}$ peak energy traces the doping dependence of $T_c$.[8] The proposed interpretation of this behavior[9] is based on the assumption of a single gap together with a change in the effective Fermi surface, which is different from the electronic picture deduced from ARPES.[6] It is also a puzzle that the $B_{1g}$ Raman peak energy is observed as always smaller than the ARPES antinodal gap energy.[10] Although some research groups reported that these two are the same,[8,11] we need to compare the two measurement (ARPES and Raman) results for the same sample to draw a definite conclusion.

The questions can be summarized as follows. (i) Can the doping dependence of the $B_{2g}$ Raman spectra be explained by the ARPES data that show a doping-independent $d$-wave gap slope along the Fermi surface? (ii) What is the $B_{1g}$ Raman gap or ARPES gap in the antinodal region, and what is the origin of its unusual doping dependence? (iii) Is the $B_{1g}$ Raman gap really smaller than the ARPES antinodal gap? If so, what is the reason? The purpose of this study is to answer these questions and to draw a unified electronic picture that reconciles the ERS and ARPES observations.

So far, there have been some reports on the calculation of ERS spectra from ARPES data. The kinetic theory calculation successfully proved that a $d$-wave pairing symmetry can explain the ERS of the cuprates.[12] However, the calculated spectra are too simple to compare with experimental data. Even though a scattering rate is introduced to explain a real broad peak, the theory can neither explain the $B_{1g}$ peak energy in the underdoped regime nor the doping dependence of the $B_{2g}$ peak energy. In order to explain the different behaviors of $B_{1g}$ and $B_{2g}$ spectra, a parameter called the quasiparticle spectral weight was added to the kinetic

theory.[8,9,11]) However, it is difficult to quantitatively compare the largely deformed spectra of the kinetic theory with experimental spectra and choose a proper scenario from the proposed candidates.[9])

Recently, it was demonstrated that ERS calculated from ARPES data through self-energy functions shows good agreement with the experimental ERS for overdoped samples in the normal state.[13]) However, the calculations of ERS in the underdoped regime and in the superconducting state were not satisfactory, which led the authors to conclude that the $B_{1g}$ Raman spectra do not represent the maximum gap that was detected by ARPES. In this paper, we extend this approach to calculate ERS in the superconducting state using the Kubo formula and the ARPES data over the entire BZ with a Shirley background subtraction. For a quantitative comparison of ERS and ARPES spectra, it is crucial to measure both spectra on the same sample. In the present study, $Bi_2Sr_2CaCu_2O_{8+\delta}$ (Bi2212) crystals at three doping levels were prepared and their ERS and ARPES were studied in order to answer the questions mentioned above and to construct a unified picture for the superconducting gap in the cuprates.

## 2. Experiments

Bi2212 single crystals were grown by a floating zone method. The carrier doping level was controlled by post-annealing under various conditions. We prepared samples at three doping levels: underdoped ($T_c$ = 75 K), nearly optimally doped ($T_c$ = 92 K), and overdoped ($T_c$ = 85 K). Hereafter, we call them UD75K, OP92K, and OD85K, respectively.

ARPES and ERS measurements were carried out in the superconducting state at 10 K. The ARPES for the samples OP92K and UD75K were measured at Stanford Synchrotron Radiation Lightsource beamline 5-4 using 22.7 eV photons with an energy resolution of 5 meV and an angular resolution of 0.1°. The polarization was along the Cu-O direction. The ARPES data of these samples were reported in Refs.7 and 14. The ARPES of sample OD85K was measured at the Institute for Solid State Physics, Tokyo University, using helium lamp light with a photon energy of 21.2 eV without polarization. The energy resolution was 10 meV, and the angular resolution was 0.1°. The energy dispersion curve (EDC) spectra were divided by the Fermi-Dirac function convoluted by the energy resolution to wipe out the thermal effect. Figure 1 shows the Fermi surface mapping images and the gap profiles along the Fermi surfaces.

ERS measurements were performed in $B_{1g}$ and $B_{2g}$ geometries on the samples from the same batch with a triple-grating Jobin-Yvon T64000 spectrometer and an Ar-Kr laser line (514.5 nm). The laser power was kept at ~ 5 mW to avoid overheating. The $B_{1g}$ geometry is obtained when crossed polarizations for incident and scattered light are rotated 45° from the Cu-O bond directions, while $B_{2g}$ polarizations are along them. In these geometries, it is possible to probe the antinodal and nodal regions corresponding to the principal axes and the diagonal of the BZ, respectively. All Raman spectra were corrected by the instrumental spectral response and the Bose factor.

## 3. Calculation method

ARPES intensity $I_{k,\omega}$ is a function of matrix elements $M_k$, Fermi Dirac function $f_\omega$, and a spectral function $A_{k,\omega}$:[5]

$$I_{k,\omega} = I_0 \cdot M_k \cdot f_\omega \cdot A_{k,\omega}. \quad (1)$$

If the matrix elements do not have a strong momentum dependence, the spectral function $A_{k,\omega}$ can be obtained directly from the ARPES spectra.

On the other hand, electronic Raman response $\chi''$ in the superconducting state can be described by Green's functions using the Kubo susceptibility[4,15] as follows

$$\chi''_{\gamma\Gamma} = \frac{2}{\pi V} \sum_k \gamma_k \Gamma_k \times \int_{-\infty}^{\infty} (f_\omega - f_{\omega+\Omega}) \times G''_{k,\omega} G''_{k,\omega+\Omega} \left(1 - \frac{\Delta_k^2}{(\omega+\zeta_k)(\omega+\Omega+\zeta_k)}\right) d\omega, \quad (2)$$

where $\Omega$ is the Raman shift, and $\gamma_k$ and $\Gamma_k$ are the bare and renormalized Raman vertices, respectively, $\Delta_k$ is the superconducting gap, $\zeta_k$ is the bare band energy, V is the volume, and $f_\omega$ is the Fermi function. Since Green's functions are related to the spectral functions through the following relation[4]

$$G''_{k,\omega} = -\pi A_{k,\omega}, \quad (3)$$

the electronic Raman responses can be calculated from the ARPES spectra. Here, we normalized the ARPES spectra for different momentum cuts by the maximum values of the energy dispersion curves at the Fermi vector. In this study, we used the same Raman vertex for both the bare and renormalized vertices. A tight binding model is often used for the band structure in cuprates, which allows us to obtain the Raman vertices[4,12] and describe the experimental ARPES data. The tight binding model with $t$ and $t'$ hopping limitations is

$$\zeta_k = -2t(\cos k_x a + \cos k_y a) + 4t' \cos k_x a \cdot \cos k_y a - \mu, \quad (4)$$

where $t$ and $t'$ are the nearest-neighbor- and next-nearest-neighbor-hopping integrals, respectively, and $\mu$ is a chemical potential. The Raman vertices for $B_{1g}$ and $B_{2g}$ geometries for

a tetragonal structure with a lattice constant $a$ and electron mass $m$ are

$$\gamma_{B_{1g},k} = \Gamma_{B_{1g},k} = ma^2 t(\cos k_x a - \cos k_y a), \quad (5)$$

$$\gamma_{B_{2g},k} = \Gamma_{B_{2g},k} = 4ma^2 t' \sin k_x a \cdot \sin k_y a. \quad (6)$$

The parameters $t$, $t'$, and $\mu$ were obtained by fitting the ARPES data with the tight binding band. For sample OP92K, $t = 0.238$ eV, $t' = 0.392t$, and $\mu = -0.320$ eV. These parameters are consistent with those of previous reports.[13] For sample UD75K, $t = 0.180$ eV, $t' = 0.490t$, and $\mu = -0.200$ eV. For sample OD85K, $t = 0.155$ eV, $t' = 0.440t$, and $\mu = -0.190$ eV.

In these calculations, the density of states over the entire BZ were taken into account, and the unoccupied states in the superconducting state were obtained by symmetrizing the ARPES EDCs against the Fermi level. A symmetric behavior of the density of state between the occupied and unoccupied state is supported by the tunneling spectra of Bi2212 when both states are probed.[16]

As is well known, in photoemission experiments, a large contribution of secondary electrons results in an intrinsic background in the spectra. This should be concerned, when we determine exact peak positions and spectral weights. In this study, a phenomenological background called the Shirley background[17] was subtracted from the raw ARPES EDC spectra. One example of this subtraction is presented in Fig. 2 (a).

For sample OD85K, we had to take into account the bilayer splitting into a bonding band (BB) and antibonding band (AB). In this study, EDC spectra were fitted by three Gaussian functions corresponding to a BB-peak, an AB-peak, and an incoherent part, as shown in Fig. 2 (b). The calculation of ERS was done for BB and AB separately. Finally, the spectra were summed to obtain the ERS spectrum for OD85K. Since the contribution from AB and BB are different, in particular, in higher energy ERS, the calculation by considering both bands gives a better fit than that from only one of the bands.

## 4. Results and discussion

The $B_{1g}$ and $B_{2g}$ Raman spectra calculated from the ARPES data are presented in Fig. 3, together with the experimental spectra. The peak intensities of the calculated and experimental spectra were normalized at the maximum. The experimental data are in good agreement with those of previous reports.[9] Roughly speaking, the overall spectral features are well reproduced by the calculation from the ARPES data. This is significant if you compare the present Kubo formula calculation to the calculation based on the kinetic theory with only $k_F$ states (dashed curves in Fig. 3).[18] Despite using the same ARPES data, the kinetic theory calculation gives very narrow $B_{1g}$ peaks with the peak energies shifted from the experimental data because delta-function-like intensity only at $k_F$ was used. Therefore, the Kubo formula calculation presented in Sec. 3 is clearly advantageous to describing ERS. The use of ARPES data over the entire BZ naturally introduces a scattering rate effect, giving a more realistic peak profile. The gap excitations at $k$ other than $k_F$ are involved, which increases the higher energy intensity of ERS, in particular that of $B_{2g}$ ERS. This is because in the nodal region, the bands near the Fermi level are quite dispersive and thus the higher energy excitation contributes to ERS.

When the carrier doping is reduced, the $B_{1g}$ peak energy shifts to higher energy in both the experimental and calculated spectra. It should be noted that the $B_{1g}$ peak energy is always lower in the experiment than in the calculation from ARPES, and this difference becomes larger with underdoping. For the $B_{2g}$ spectra, the difference between the experimental and calculated spectra is remarkable in the low-energy region, and it becomes larger with underdoping.

Next, to further improve our calculation, we considered the effect of matrix element $M_k$, which was assumed to be constant in the calculations so far. In reality, $M_k$ is not uniform in BZ but changes with $k$ depending on polarization and measurement geometry. This modifies the ARPES spectral intensity $I_{k,\omega}$. This means that the real intensity distribution of $A_{k,\omega}$ along the Fermi surface cannot be easily deduced from the ARPES intensity $I_{k,\omega}$. Since it is difficult to determine $M_k$ experimentally, we assume here a simple linear peak-intensity profile of $A_{k,\omega}$ along the Fermi surface. In fact, a linear behavior of the intensity profile was found in the ARPES data of $(La,Sr)_2CuO_4$,[19] and thus it is reasonable to assume such a linear profile.

Figures 4 (a)-(c) present the trial profiles. The intensity in the nodal region was fixed and reduced/increased toward the antinodal region. Here, for example, N10-AN1 denotes an intensity that is 10 times larger in the nodal spectra than in the antinodal ones. The EDC spectra on different cuts of ARPES measurements were multiplied by corresponding factors that converted an experimental profile along the Fermi surface to an assumed profile. The $B_{1g}$ and $B_{2g}$ ERSs were then calculated from these ARPES EDCs, as demonstrated in Figs. 4 (d)-(i). We see that even if we vary the intensity profiles, there is no significant change in the $B_{1g}$ peak position, while the $B_{2g}$ spectra are improved.

For sample OD85K, a good fit is obtained with profile N10-AN10 that has a constant intensity along the Fermi surface. For sample OP92K, the profile N10-AN5 gives the best fit. The reduction of intensity in the antinodal region pushes up the low-energy part in the $B_{2g}$ spectrum. For sample UD75K, we need a more radical reduction in the antinodal region to obtain a good agreement between the calculated and experimental spectra. When the profile changes from N10-AN10 to N10-AN1, we find a tendency that the spectral weight in the low-energy region of both $B_{1g}$ and $B_{2g}$ is increased and the $B_{2g}$ peak shifts to lower energy. We also examined a special profile in which the intensity changes more radically, as indicated in Fig. 4 (c). This special profile gives a better fit in the $B_{2g}$ spectra.

Summarizing the calculation of the Raman spectra with the assumed intensity profiles of $A_{k,\omega}$, we find that different doping levels have different intensity profiles: a special profile for the underdoped UD75K, N10-AN5 for the optimally doped OP92K, and N10-AN10 for the overdoped OD85K. It is difficult to obtain this doping dependence of the spectral weight distribution from ARPES measurements because the matrix elements change the intensity profiles of the spectral functions. In this sense, the present trial functions remove the matrix element effects. (See the Appendix.) The result shows that $A_{k,\omega}$ confined in the nodal region distributes to the antinodal region as the doping level increases. This behavior is consistent with a recent report that the superconducting spectral weight is suppressed by the competing pseudogap in the antinodal region.[20] Therefore, the behavior can be understood as the recovery of the superconducting spectral weight with doping as a result of the weakening of the pseudogap.

Note that in the special profile of sample UD75K, the portion of the Fermi surface that has a strong intensity is up to 0.4 on the horizontal axis in Fig. 4 (c). This Fermi arc should contribute to superconductivity. However, the Fermi arc region that is gapless in the normal state as defined by ARPES for this sample is just up to ~0.25.[14] Therefore, the ARPES Fermi arc may not represent the real superconducting region, but a larger Fermi surface region contributes to superconductivity.

From Fig. 4, it is clear that the $B_{2g}$ spectra are well reproduced by the ARPES data assuming proper peak-intensity profiles of $A_{k,\omega}$ along the Fermi surfaces. This implies that the apparent peak shift of the $B_{2g}$ Raman spectra with doping can be understood by the superconducting gap profile of ARPES that has a doping-independent nodal slope.[6,7] Therefore, the observed peak shift of the $B_{2g}$ spectrum does not necessarily indicate a decrease in the superconducting gap.

The peak energies of the calculated Raman spectra are plotted in Fig. 5 together with the experimental Raman, ARPES, and STM data.[7,22] Here, the published ARPES data[7] were taken for samples from the same batch as ours. The calculated $B_{1g}$ and $B_{2g}$ peak energies are close together in the overdoped sample but become separated with underdoping, as reported previously.[8,9] Note that the ARPES antinodal gap is always larger than the experimental $B_{1g}$ peak energy, and the difference increases with underdoping.

One can also notice that the calculated $B_{1g}$ peak energy is closer to the ARPES antinodal gap rather than the experimental $B_{1g}$ energy. The difference from the experimental Raman data increases with underdoping. The fact that the $B_{1g}$ peak energy calculated from the ARPES data reproduces the ARPES antinodal gap energy indicates that our calculation method is appropriate. As demonstrated previously, the intensity profile of $A_{k,\omega}$ cannot cause a shift in the $B_{1g}$ position. Therefore, this difference between the experimental Raman $B_{1g}$ and the ARPES antinodal gap energies must be intrinsic.

Since the difference increases with underdoping, this difference is possibly caused by the pseudogap. In Fig. 5, we plotted the pseudogap energy determined by ARPES at 100 K.[7] The pseudogap increases rapidly with underdoping and it seems that the superconducting gap in ARPES is enhanced by an underlying high-energy pseudogap. Here, we recall that the gap profile of sample UD75K deviates from *d*-wave behavior in the antinodal region (Fig. 1(d)), which can be caused by the pseudogap. The unusual doping dependence of the Raman $B_{1g}$ gap energy, namely, the monotonous increase with underdoping, can also be considered an effect of the pseudogap. Moreover, the difference between the Raman $B_{1g}$ and the ARPES antinodal gap energies indicates that the effect of the pseudogap manifests itself differently in different measurement techniques. The effect on ARPES is stronger than on Raman. It is also well known that in experiments, the pseudogap is only weakly visible in Raman (giving a weak suppression of low-energy intensity) but is clear in ARPES data (giving a peak or shoulder).

## 5. Conclusion

In this study, two *k*-selective measurements (ARPES and ERS) were performed for the same Bi2212 single crystals with different doping levels at 10 K, well below $T_c$. The $B_{1g}$ and $B_{2g}$ Raman spectra were calculated from the ARPES data over the entire Brillouin zone by using the Kubo formula and assuming the *k*-dependence of the peak-intensity profile of the spectral function $A_{k,\omega}$. The calculated Raman spectra well reproduced the experimental data except for the $B_{1g}$ peak energies. This indicates that we successfully established an analytical method by which to compare ARPES and Raman data.

From the present results, we reached the following conclusions. First, Raman and ARPES can be understood with the same gap profile. Namely, the nodal slope of the gap profiles is doping independent, as reported by ARPES. The apparent doping dependence of the $B_{2g}$ peak energy is caused by the change in spectral weight of $A_{k,\omega}$ along the Fermi surface. Second, the antinodal gap of ARPES is a superconducting gap that is strongly affected by the pseudogap, whereas the Raman $B_{1g}$ gap is moderately affected. This probe-dependent effect of the pseudogap is the main source for the difference between the Raman $B_{1g}$ gap and the ARPES antinodal gap energies. Third, while the spectral weight of $A_{k,\omega}$ is confined to the nodal region in the underdoped sample, the antinodal region gains spectral weight with doping and contributes to superconductivity. Although this is similar to the "Fermi arc" picture reported previously,[14] the Fermi surface area contributing to superconductivity is larger than that estimated from the normal state ARPES as a Fermi arc. All of these findings reflect the unusual electronic states where superconductivity and pseudogaps coexist even at the lowest temperatures.

**Acknowledgment**

The authors thank Prof. T. P. Devereaux for his useful discussion. They also thank Prof. S. Shin and Dr. Y. Ishida for their help in ARPES measurements. This work was supported by a Grant-in-Aid for Scientific Research (B) from JSPS (No. 24340083).

**Appendix**

From the assumed intensity profile of $A_{k,\omega}$ and the experimentally observed intensity $I_{k,\omega}$ along the Fermi surface, the matrix element $M_k$ can be estimated as shown in Fig. A1. This cannot be obtained from the ARPES data directly but is first extracted in the present Raman fitting procedure. Despite our oversimplified profiles of $A_{k,\omega}$ in Fig. 4, the obtained $M_k$ is roughly monotonically $k$-dependent where $M_k$ is small in the nodal region and large in the antinodal region. The approximately smooth curves of $M_k$ indicate that our selection of $A_{k,\omega}$ profile is appropriate.

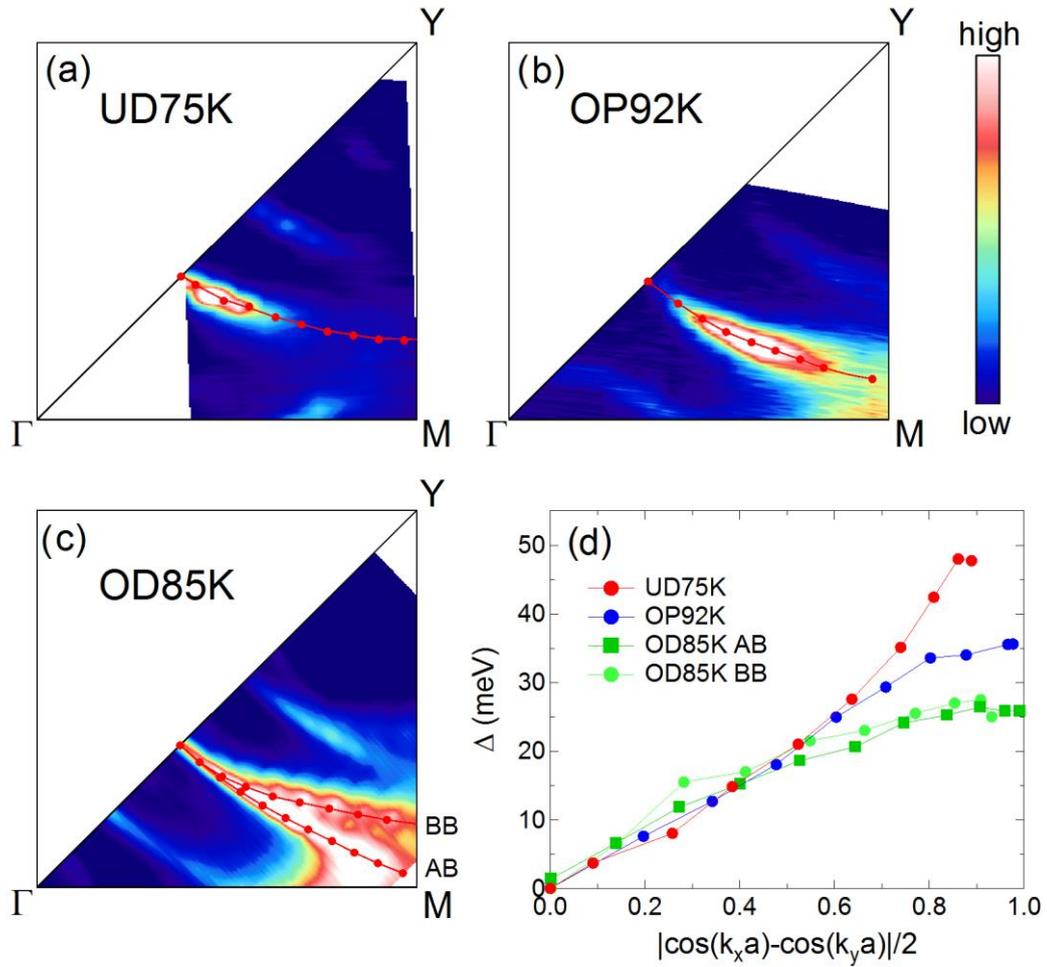

Fig. 1. (a)-(c) ARPES Fermi surface mapping images of three Bi2212 samples. Red dots are $k$-points on Fermi surface. $k_F$ point at node was added to see full Fermi surface. For sample OD85K, there is a split into an antibonding band (AB) and bonding band (BB). (d) Gap profiles along Fermi surfaces against $d$-wave function.

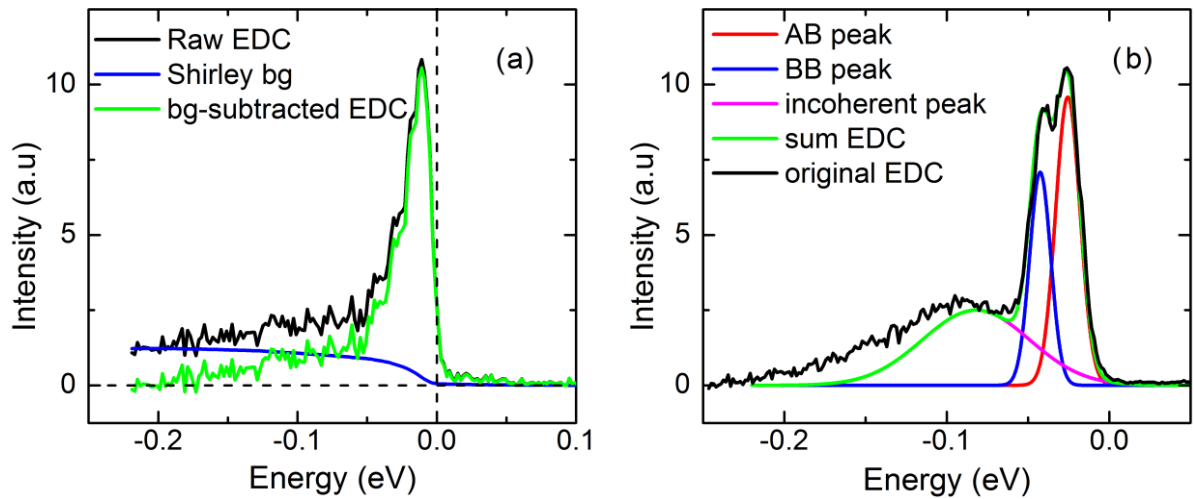

Fig. 2. (a) Shirley background (bg) subtraction for ARPES EDC spectra, with one EDC spectrum on Fermi surface of sample OP92K. (b) Subtraction for AB band from ARPES data of sample OD85K.

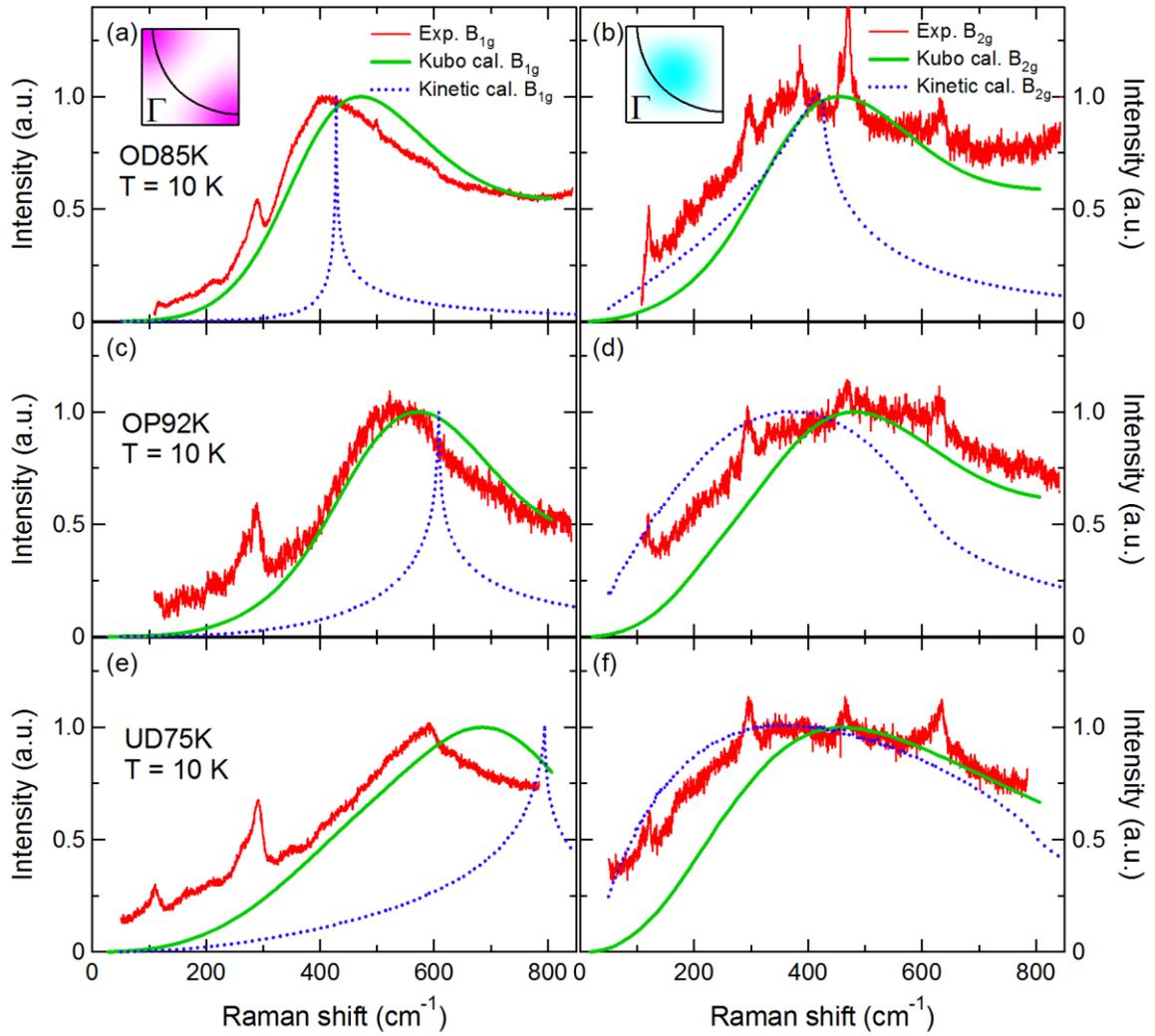

Fig. 3. $B_{1g}$ [(a), (c), (e)] and $B_{2g}$ [(b), (d), (f)] ERS spectra of Bi2212. Calculations using Kubo formula and ARPES data over entire BZ after Shirley-background subtraction (Kubo cal. $B_{1g}$ and Kubo cal. $B_{2g}$) are compared with experimental data (Exp. $B_{1g}$ and Exp. $B_{2g}$). Calculation results from kinetic theory (Kinetic cal. $B_{1g}$ and Kinetic cal. $B_{2g}$) using ARPES data are indicated by dashed curves.

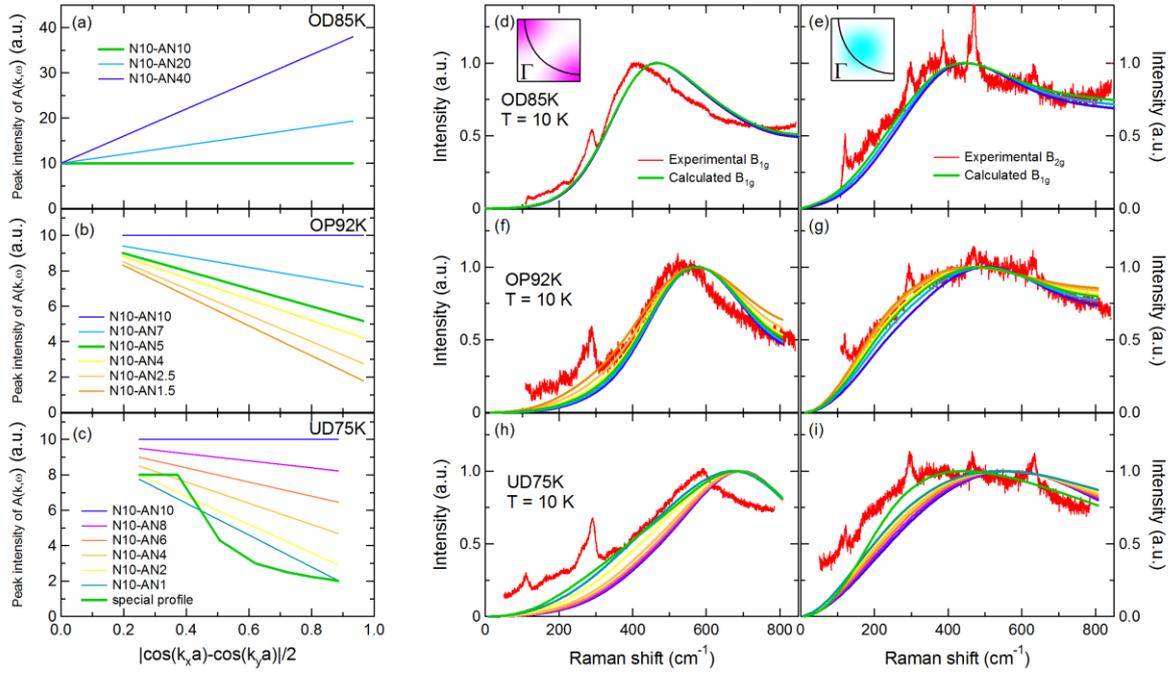

Fig. 4. (a)-(c) Different assumed peak-intensity profiles of spectral functions $A_{k,\omega}$ for three Bi2212 samples. (d)-(i) ERS $B_{1g}$ and $B_{2g}$ calculation results, in which profiles were applied, in comparison to experimental data. Best profiles for samples OD85K, OP92K, and UD75K are N10-AN10, N10-AN5, and special profile, respectively.

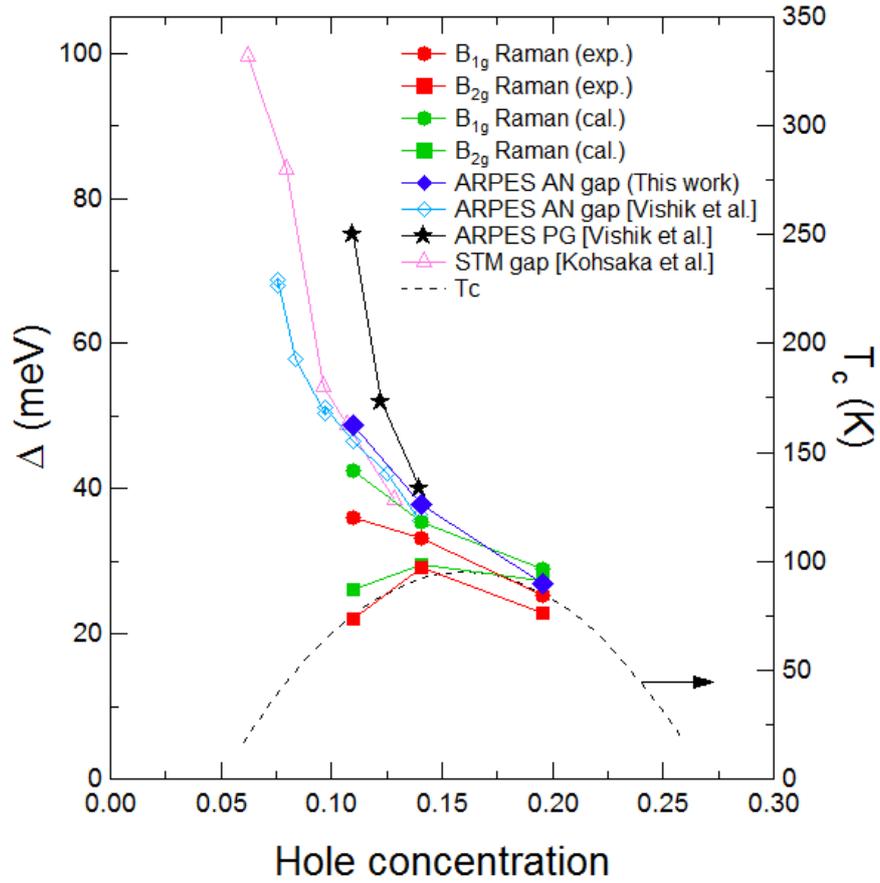

Fig. 5. Doping dependence of peak energies in Bi2212 obtained from ERS calculations in comparison with experimental data from Raman, ARPES, and STM measurements. AN: antinodal, PG: pseudogap. Dashed curve is $T_c$ dome obtained from empirical equation $T_c = T_{\max c}[1 - 82.6(p - 0.16)^2]$ [21]) with $T_{\max c}$ = 93 K. Error bars are 2 meV.

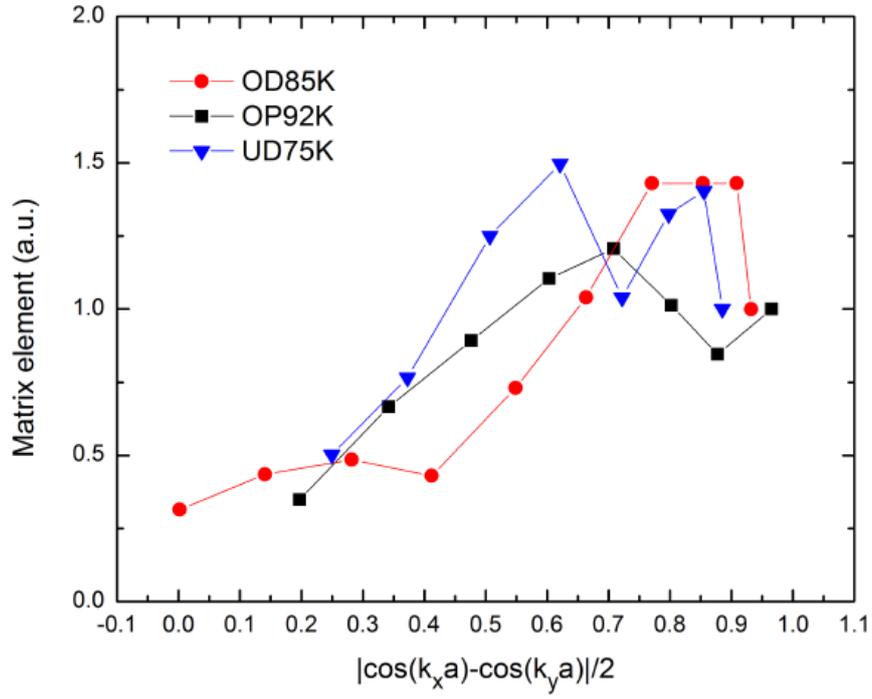

Fig. A1. Matrix element $M_k$ estimated from assumed intensity profile $A_{k,\omega}$, and experimental $I_{k,\omega}$ along Fermi surface (normalized for values in antinodal region).